\documentclass[prl,twocolumn,showpacs,floatfix,superscriptaddress]{revtex4}
\usepackage{amssymb}
\usepackage{epsfig}

\begin{document}


\title{Designing $\pi$--conjugated polymers with light emission in the infrared}
\author{S. Mazumdar}
\affiliation{ Department of Physics, University of Arizona
Tucson, AZ 85721}
\author{S. Dallakyan}
\affiliation{ Department of Physics, University of Arizona
Tucson, AZ 85721}
\author{M. Chandross}
\affiliation{Sandia National Laboratories, Albuquerque, NM 87185-1411}

\begin{abstract}
There is currently a great need for solid state lasers that emit in the
infrared.
Whether or not conjugated polymers that emit in the IR can be synthesized is an
interesting theoretical challenge. We show that emission in the IR can be
achieved in designer polymers in which the effective Coulomb correlation is smaller
than that in existing systems. We also show that the structural requirement for having
small effective Coulomb correlations is that there exist transverse $\pi$--conjugation
over a few bonds in addition to longitudinal
conjugation with large conjugation lengths.
\end{abstract}
\maketitle

All light emitting $\pi$-conjugated polymers
to date emit in the visible or UV. Telecommunications use infrared
radiation, so lasing at these wavelengths is desirable. Within conventional
theories of light emission from $\pi$-conjugated polymers,  
light emission in the IR from undoped $\pi$-conjugated polymers would be 
impossible. In this paper we point out the structural modifications that
can lead to emission in the IR.

Linear polyenes and trans-polyacetylene (t-PA) are
weakly emissive, because the lowest two-photon
state, the 2A$_g$, occurs below the optical 1B$_u$.
The optically pumped 1B$_u$ decays to the 2A$_g$ in ultrafast times, and
radiative transition from the 2A$_g$ to the ground state 1A$_g$ is forbidden.
Strong photoluminescence (PL) in systems like PPV and PPP implies 
E(2A$_g$) $>$ E(1B$_u$) (where E(...) is the energy of
the state). This reversed excited state ordering can be understood 
by enhanced bond alternation
within the effective linear chain model for
PPV, PPP, etc. \cite{Soos}. 
Since enhanced bond alternation necessarily
{\it increases} E(1B$_u$), it appears that strong PL should be limited
to systems with optical gaps larger than that of t-PA.

Our goal is to demonstrate theoretically that materials obtained by
``site-substitution'' of t-PA, in which the hydrogen atoms of t-PA are replaced
with transverse conjugated groups 
of finite size 
will 
have small optical
gaps {\it and} E(2A$_g$) $>$ E(1B$_u$). 
Systems obtained by such site-substitution consist of  
molecular units linked by
a single longitudinal bond. With moderate e-e interactions, the ground state
can be thought of as covalent 
(all atomic sites singly occupied). 
Optical
excitation along the backbone chain
involves inter-unit one-electron hops that
generate C$^+$ and C$^-$ ions on different units. These charges 
are delocalized over 
entire molecular units, leading to considerable decrease in the 
{\it effective}
on-site Coulomb interaction 
$U_{eff}$. Smaller $U_{eff}$
can now 
give smaller E(1B$_u$) (relative to t-PA) and
E(2A$_g$) $>$ E(1B$_u$).

Initial work along this direction
by us was limited to 
poly-diphenylpolyacetylene (PDPA) \cite{Shukla,Ghosh}, in which the 
substituents are phenyl groups.
PDPAs have short conjugation lengths \cite{Tada} and
emit in the visible. Further, although we determined
that E(2A$_g$) $>$ E(1B$_u$) in PDPA \cite{Shukla,Ghosh},
because of the  
complexity of its molecular structure, it is not ideal
for illustrating a general theoretical principle. 
We consider here
the simplest possible $\pi$-conjugated polymer with conjugated sidegroups, the
hypothetical 
polymer shown in Fig.~1,
in which the substituents are ethylenes.
This simple structure allows more insight 
than more complicated ones. 
\begin{figure}
\begin{center}
\epsfysize=4cm\epsfbox{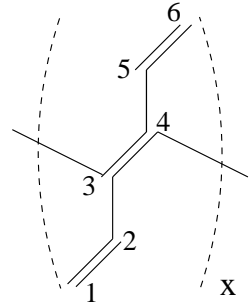}
\caption{The hypothetical polymer investigated theoretically}
\end{center}
\end{figure}

We consider linear polyenes as well as 
oligomers of the structure in Fig.~1 
within the dimerized Hubbard
Hamiltonian,
\begin{eqnarray}
H = H_{1e} + H_{ee} \\
H_{1e} = -\sum_{\langle ij \rangle,\sigma}t_{ij}c_{i,\sigma}^\dagger c_{j,\sigma} \\
H_{ee} = U\sum_{i}n_{i,\uparrow}n_{i,\downarrow}
\label{Hubbard}
\end{eqnarray}
\noindent In the above, $H_{1e}$ describes one-electron nearest neighbor
hops of electrons, and $H_{ee}$ is the electron-electron
(e-e) interaction.
We have used
$t_{ij}$ = 2.2 eV and 2.6 eV for
single and double bonds, respectively. 
For both the unsubstituted and the substituted
polyene we increase 
$U$ from zero (where E(2A$_g$) $>$ E(1B$_u$))
and determine the critical $U_c$ at which 
E(2A$_g$) $<$ E(1B$_u$) occurs. A smaller $U_{eff}$ in the substituted polymer
requires $U_c$(substituted) $>$ $U_c$(unsubstituted). 

The complete Hamiltonian is way beyond our reach for 
more than two units of the substituted polyene of
Fig. 1. We thererfore choose an appropriate minimal basis for the substituted system
that allows many--body accurate
calculations. 
Our approach is based on the
exciton basis valence bond method \cite{Chandross}, within which a linear polyene
is considered as coupled ethylenic dimer units, and the polymer of Fig.~1 would
be coupled hexatrienes. The basis functions are valence bond
(VB) diagrams with bonds between the molecular orbitals (MOs) of the individual units.
We write
$H_{1e} = H^{intra}_{1e} + H^{inter}_{1e}$, where the two terms are the intra- and 
inter-unit part of $H_{1e}$. We solve $H^{intra}_{1e}$ exactly, and
rewrite the Hamiltonian of Eq.~(1) in terms of a new basis set that consists of the 
MOs of individual hexatriene units. The minimal basis set  
now consists of the HOMO and the LUMO of the molecular hexatriene unit. In Fig.~2 we have
shown the one--electron energies of the MOs of hexatriene with their energies in eV.
The relatively large energy gaps involving the higher antibonding MOs and lower bonding
MOs partly justify the minimal basis. Furthermore, 
$H_{ee}$ within Eq.(1) creates only intraunit excitations
\cite{Chandross}, in contrast to the standard MO-based CI approaches, within which the
Hubbard interaction is nonlocal, thereby making the choice of a minimal basis far more 
difficult within the standard approaches. 
Additional justification for the minimal basis
will be presented later.

\begin{figure}
\begin{center}
\epsfysize=4cm\epsfbox{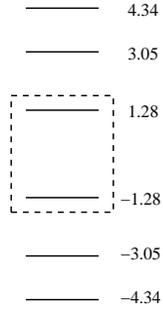}
\caption{The single-particle energies (in eV) for each isolated unit,
with standard hopping integrals (see text). The dashed box indicates our
minimal basis.}
\label{orbitals}
\end{center}
\end{figure}

We give here a brief introduction to the 
exciton basis for polyenes \cite{Chandross}. The fundamental zero, one
and two-excitations for butadiene (two coupled ethylenes) 
are shown in Fig.~3. The leftmost
diagram in Fig~3(a) is the 
``ground state'' zero-excitation diagram, with all electrons in the bonding
MOs of the two dimer units. Application of $H_{inter}^{1e}$ on this gives the 
two charge--transfer (CT) VB diagrams with singlet bonds between the bonding MO
of one unit and the antibonding MO of the other unit (thus each arrow in Fig.~3 
indicates one application of $H_{inter}^{1e}$), while a second 
application of $H_{inter}^{1e}$ generates the VB diagram with crossed bonds, 
the rightmost diagram in Fig.~3(a).
The CT diagrams are coupled to intra-unit one-excitations
by $H_{inter}^{1e}$, as is shown in Fig.~3(b), while the crossed VB diagram
is a superposition of two other two-excitations with noncrossing singlet bonds,
as is shown in Fig~3(c). The particular 2:1 superposition of these two
VB diagrams is a triplet--triplet (TT) excitation, with two triplets on the
two dimer units \cite{Chandross}, as shown in Fig.~3(d). 
The diagrams in 
longer chains are similar to those shown
in Fig.~3, -- the only difference arises in the lengths of the interunit
singlet bonds, which can be between nonnearest neighbor units in long chains. 
Furthermore, in the minimal basis (see Fig.~2), the same VB diagrams occur
in the oligomer of Fig.~1 as in a linear polyene of the same length, the only
difference arises from the fact that
the expansions of the HOMO and the LUMO in the
substituted polyene in terms of the atomic orbitals are different
from that in the unsubstitbuted polyene, and hence  
the matrix elements of $H_{ee}$ and 
$H_{inter}^{1e}$ are different.

\begin{figure}
\begin{center}
\epsfysize=3.5cm\epsfbox{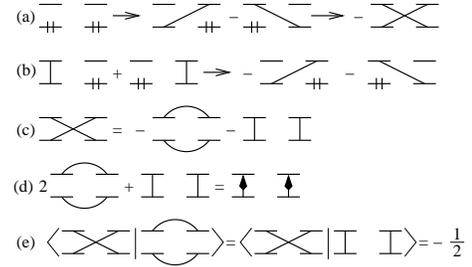}
\caption{Typical exciton basis VB diagrams for butadiene or the two-unit oligomer of
Fig.~1 within the minimal basis. Each molecule is considered as coupled dimer units,
and electrons occupy the bonding and antibonding MOs of the dimers. Singly
occupied MOs are connected by spin singlet bonds. The diagram on the right hand
side of (d) denotes two triplet excitations. Relationship (e) shows that exciton
VB diagrams are nonorthogonal.}
\end{center}
\end{figure}

We performed exact numerical calculations for unsubstituted and
minimal basis substituted polyenes with N = 4, 6 and 8 backbone
carbon atoms and quadruple-CI (QCI) calculations for the N = 10 system
to determine $U_c$ for each of these systems. 
QCI gives practically exact energies for the 
lowest eigenstates
at these chain lengths \cite{Tavan,FGuo}. In Fig.~4 we have plotted
the excitation energies E(1B$_u$) and E(2A$_g$) [E(1A$_g$) = 0] for the N = 10
cases. Note that not
only is the E(1B$_u$) for the substituted system smaller, E(1B$_u$) in this
case increases with $U$ much less rapidly than for the unsubstituted polyene.
More importantly, the $U_c$ for the substituted system is larger than that
for the unsubstituted polyene by more than 1 eV. 
The initial increase in E(2A$_g$) for the 
substituted polyene is real and has been seen previously in 
calculations for long polyenes \cite{Shuai}, where it was
found that the smaller the 2A$_g$ -- 1B$_u$ gap at $U$ = 0, the more the
tendency to initial increase of E(2A$_g$). This particular energy gap is
obviously smaller in the substituted polyene. 
\begin{figure}
\begin{center}
\epsfysize=4cm\epsfbox{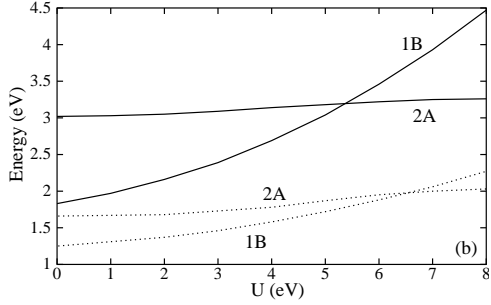}
\caption{E(1B$_u$) and E(2A$_g$) for
the unsubstituted (solid lines) and the substituted (dotted lines) polyene, for N = 10
backbone carbon atoms.}
\end{center}
\end{figure}

Fig.~5 shows the difference $\Delta U_c$ between the $U_c$ values
for the substituted and the unsubstituted polyenes against 1/N. $\Delta U_c$
increases modestly between N = 4 and 8, but then appears to saturate. The
tendency to saturation in $\Delta U_c$ is an indirect signature of the 
accuracy of the minimal basis calculations: since
the number of levels being ignored in the incomplete CI calculations for the
susbstituted cases increases with N, had the larger $U_c$ for the 
substituted polyenes been a consequence of incomplete CI, $\Delta U_c$
would have increased continuously with increasing N.

\begin{figure}
\begin{center}
\epsfysize=4cm\epsfbox{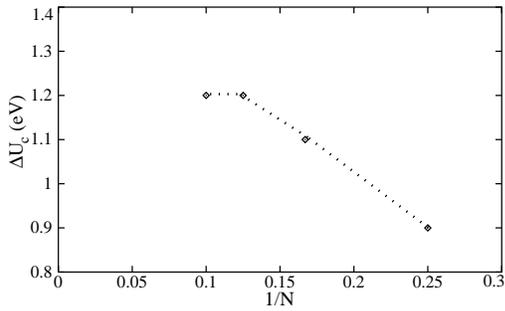}
\caption{The difference in $U_c$ between substituted and unsubstituted
polyene as a function of $1/N$.}
\end{center}
\end{figure}

In addition to calculations of energies, we have also done wavefunction
analysis of the 1B$_u$ and the 2A$_g$ to verify that the effective on-site
Coulomb correlation is smaller in the substituted polyene. While the 
1B$_u$ wavefunctions are very similar at fixed $U$, 
the 2A$_g$ wavefunctions of the unsubstituted and the substituted polyenes are 
substantially different. For the specific case of $U$ = 6 eV in Fig.~4,
the 2A$_g$ of the polyene is
composed almost entirely of TT two-excitations (see Fig.~3(d)), while the largest
contributions to the 2A$_g$ of the substituted polyene come from the CT
one-excitations (see Figs.~3(a) and (b)) \cite{Dallakyan}. Indeed, the 2A$_g$ of the 
substituted polyene is very similar to the 2A$_g$ of the unsubstituted polyene
with large bond-alternation \cite{Chandross}, for which also 
E(2A$_g$) $>$ E(1B$_u$). We thus see that at least within the minimal basis, the substituted
polyene has a substantially smaller $U_{eff}$.

We now argue that complete CI, viz., going beyond the minimal basis of 
Fig.~2 will not alter the basic result that there exists a range of $U$ over which
the E(2A$_g$) $>$ E(1B$_u$) in the substituted polyene but
E(2A$_g$) $<$ E(1B$_u$) 
in the unsubstituted material. CI between the minimal basis TT exciton basis VB diagrams 
that have been retained in our
calculations and higher energy TT diagrams
that have been ignored can potentially lower the 2A$_g$ energy in the substituted
polyene.
The higher energy TT diagrams are
of two kinds, (i) TT excitations involving a single molecular unit, and (ii) those
involving two units. We first discuss the role of type 
(i) TT excitations. In Fig.~6(a) - (d)we have shown four of the basis functions that dominate
the molecular 2A$_g$ state \cite{Hudson}
(here the lowest three MOs are bonding, the higher three
are antibonding). Diagram 6(a) is a superposition of one-excitations. Fig.~6(e)
shows the CI between one of these diagrams and a minimal basis CT diagram. Such 
CI process can only raise the energy of the 2A$_g$; hence we discuss this no further.
Diagram 6(b) has already been included in our minimal basis calculations. This leaves
diagrams 6(c) and (d). These can have CI with the minimal basis TT diagrams only
through two applications of $H_{inter}^{1e}$, as shown 
explicitly in Fig.~6(f) and (g). 
Thus the TT diagrams 6(c) and (d) do indeed
give the minimal basis TT diagrams, and this suggests that in principle CI processes
such as these can lower E(2A$_g$).

\begin{figure} 
\begin{center}
\epsfysize=6cm\epsfbox{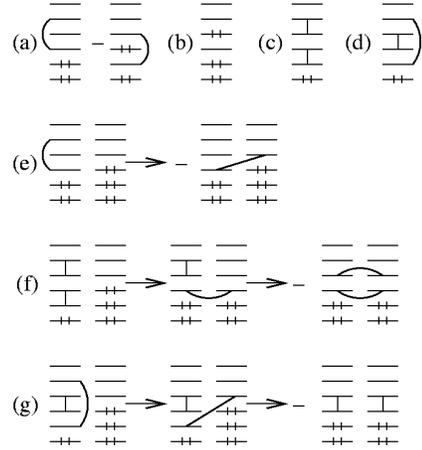}
\caption{(a) -- (d): Intraunit excitations that dominate hexatriene 2A$_g$.
(e) - (g) shows CI processes in the dimer of Fig.~1.
(e) CI process between one of the single excitations of (a) and
a minimal basis CT diagram for the dimer;
(f) and (g)  CI processes between intraunit 2e--2h excitations
and the two components of the minimal basis TT diagrams.} 
\end{center}
\end{figure}

We point out, however, that the CI processes 6(f) and (g) are {\it strictly local},
{\em i.e.}, molecular TT basis functions can be coupled only to minimal basis TT
functions in which the two triplets occur on nearest neighbor units. Since in long chains
there occur many more minimal basis TT diagrams with the triplets localized on distant
units, it is clear that CI processes 6(f) and (g) become progressively 
unimportant with increasing chain length. Equally importantly, if the molecular
unit is such that the molecular 2A$_g$ is {\it above} the molecular 1B$_u$ (this is the
case with PDPA), then CI with molecular TT functions can only raise the energy of
the polymeric 2A$_g$. Since the polymer of Fig.~1 is for the sake of illustration only,
this information can be used to further modify the basic design of the desired polymer
(for example, replacing the double bonds of the sidegroups with triplet bonds would
continue to give low E(1B$_u$) but would increase E(2A$_g$)).

CI with higher energy TT diagrams of type (ii), 
can only {\it increase} E(2A$_g$). To demonstrate this, first consider a
new restricted basis in which we retain the HOMO/LUMO (HOMO--1/LUMO+1) MOs
of odd (even) units. This system is equivalent to an ordered copolymer
in which alternate monomers have larger one--electron gaps than in ethylene
(e.g., polydiacetylene), and has higher relative E(2A$_g$) than in t-PA. 
We can now visualize many similar systems, in which we retain various pairs of
these four MOs on each unit. All such disordered copolymers also have higher
E(2A$_g$). CI with type (ii) TT diagrams effectively includes matrix elements
between A$_g$-type configurations within these new bases and the minimal basis
A$_g$ configurations, and will therefore only increase E(2A$_g$).

To summarize, transverse conjugation over a small number of bonds is one way
to generate conjugated polymers with light emission in the IR. Excited 
state ordering conducive to light emission is a consequence of
the reduction of the effective Hubbard repulsion due to delocalization over
a molecular site.
The system investigated 
here is the simplest example with this characteristic, but one can visualize
many other such structures, including realistic materials like 
poly-isothianaphthene with a gap of 1.1 eV. We hope that our work will
stimulate experimentalists to investigate derivatives of the
latter or to synthesize new materials with the necessary structural features.

Work in Arizona was supported by NSF DMR-0101659, NSF ECS-0108696 and the ONR.
Sandia is a multiprogram laboratory operated by Sandia Corporation, a
Lockheed Martin Company, for the United States Department of Energys's
National Nuclear Security Administration under
Contract DE-AC04-94AL85000.

\end{document}